\documentclass[aps,twocolumn,showpacs,amsmath,amssymb,superscriptaddress]{revtex4}

\usepackage{graphicx}
\usepackage{subfigure}
\usepackage{grffile}
\usepackage{bm}
\usepackage{color}
\usepackage{hyperref}
\usepackage{natbib}

\newcommand{\abs}[1]{\ensuremath{\left| #1 \right|}}
\newcommand{\ket}[1]{{\vert #1\rangle}}

\newcommand{\braket}[2]{\langle#1\vert#2\rangle}
\newcommand{\eval}[3]{\langle#1\vert#2\vert#3\rangle}
\newcommand{\vev}[1]{\langle #1\rangle}

\begin{document}

\title{Enhanced Charge Order in a Photoexcited One-Dimensional
  Strongly Correlated System}

\author{Hantao Lu}
\affiliation{Yukawa Institute for Theoretical Physics, Kyoto
  University, Kyoto, 606-8502, Japan}

\author{Shigetoshi Sota} 
\affiliation{Computational Materials Science Research Team, RIKEN
  AICS, Kobe, Hyogo 650-0047, Japan} 

\author{Hiroaki Matsueda}
\affiliation{Sendai National College of Technology, Sendai, 989-3128,
  Japan}

\author{Janez Bon\v{c}a}
\affiliation{Faculty of Mathematics and Physics, University of
  Ljubljana, SI-1000 Ljubljana, Slovenia}
\affiliation{J. Stefan Institute, SI-1000 Ljubljana, Slovenia}

\author{Takami Tohyama}
\affiliation{Yukawa Institute for Theoretical Physics, Kyoto
  University, Kyoto, 606-8502, Japan}

\date{\today}

\begin{abstract}

  We present a compelling response of a low-dimensional strongly
  correlated system to an external perturbation. Using the
  time-dependent Lanczos method we investigate a nonequilibrium
  evolution of the half-filled one-dimensional extended Hubbard model,
  driven by a transient laser pulse. When the system is close to the
  phase boundary, by tuning the laser frequency and strength, a
  sustainable charge order enhancement is found that is absent in the
  Mott insulating phase. We analyze the conditions and investigate
  possible mechanisms of emerging charge order enhancement. Feasible
  experimental realizations are proposed.

\end{abstract}


\maketitle

Nonequilibrium processes in strongly correlated electron systems can
provide new insights into the dynamical properties of these systems,
which, in many aspects, can be qualitatively different from their
weakly interacting counterparts. One such example is nonequilibrium
induced phase
transition~\cite{Yonemitsu:2008bz,Iwano:2004er,Iwano:2009gb}. As the
system is driven away from the equilibrium, under certain conditions,
a ``crossover'' from one state to another (metastable) state may
occur.

A well-known example is the insulator-to-metal transition induced
either by strong electric field or transient laser pulse, as a result
of
photodoping~\cite{Oka:2003hr,Oka:2005dk,Takahashi:2008bk,Ockstein2010,Okamoto:2007kc}. In
the case of Mott insulators, photocarriers are doublons and holons. In
one dimension, just after the doping, generally, the system is in a
metallic state due to the existence of itinerant carriers and the
benefit of the spin-charge
separation~\cite{Okamoto:2007kc,AlHassanieh:2008fl}. Experimentally,
the dielectric breakdown in one-dimensional (1D) Mott insulators has
been observed long ago in organic
materials~\cite{Tokura:1988,Sawano:hy} and
oxides~\cite{Taguchi:2000}. Photoinduced metallic state by transient
light has also been investigated~\cite{Iwai:2003,Okamoto:2007kc}. More
recently, the nonequilibrium and nonlinear phenomena have stimulated
significant interest within the cold atom community, where a novel
realization of the fermionic Mott insulating state has been
achieved~\cite{Jordens:2008im}.

In this Letter we address the question of whether it is possible to
reach other characteristically different states besides the metallic
one after the photodoping. One possibility is the enhancement of
charge-order when the attractive interaction between doublons and
holons is incorporated~\cite{Gomi:2005ht}. We seek such enhancement in
the extended half-filled 1D Hubbard model with an additional
nearest-neighbor interaction. At half filling, its phase is well-known
and understood~\cite{vanDongen:1994ca,
  Nakamura:2000gk,Tsuchiizu:2002eg,Ejima:2007go}. In large on-site
Coulomb interaction, the model possesses two phases: spin-density-wave
(SDW) and charge-density-wave (CDW), connected by a first order
quantum phase transition, with algebraic decay of spin correlations
and long-range charge order, respectively. In general, as a correlated
system approaches the phase boundary, the response to an external
perturbation becomes more elaborate. In this Letter we show that in
the latter case a substantial change in the electronic structure can
be triggered by the optical pulse. In particular, when the system is
originally in the Mott-insulating phase but close to the transition to
CDW, a sustainable enhancement of charge order parameter is achieved
by exposing the system to an external optical pulse with a rather
carefully tuned frequency and amplitude.

We consider the 1D extended Hubbard model at half filling. The laser
pump is incorporated by means of the Peierls substitution in the
Hamiltonian:
\begin{eqnarray}
H(t)&=&-t_h\sum_{i,\sigma}\left(e^{iA(t)}c_{i,\sigma}^{\dagger}
c_{i+1,\sigma}+\text{H.c.}\right) \nonumber \\
&+&U\sum_{i}\left(n_{i,\uparrow}-\frac{1}{2}\right)
\left(n_{i,\downarrow}-\frac{1}{2}\right) \nonumber \\
&+&V\sum_i\left(n_i-1\right)\left(n_{i+1}-1\right),
\label{eq:1}
\end{eqnarray}
where $c_{i,\sigma}^\dagger$ ($c_{i,\sigma}$) creates (annihilates)
electrons with spin $\sigma$ at site $i$,
$n_{i,\sigma}=c_{i,\sigma}^\dagger c_{i,\sigma}$,
$n_i=n_{i,\uparrow}+n_{i,\downarrow}$, $t_h$ is the hopping constant
while $U$ and $V$ are the on-site and nearest neighbor interaction
strengths, respectively. We model the external laser pulse in the
temporal gauge via the time-dependent vector potential
$A(t)$~\cite{Matsueda:2010ue}
\begin{equation}
A(t)=A_0e^{-\left(t-t_0\right)^2/2t_d^2}\cos
\left[\omega_{\text{pump}}\left(t-t_0\right)\right],
\label{eq:5}
\end{equation}
where $A_0$ controls the laser amplitude, which reaches its full
strength at $t=t_0$; $t_d$ characterizes the duration time of light
action. Notice that due to finite $t_d$, the incoming photon frequency
is broadened into a Gaussian-like distribution, with the variance of
$1/t_d^2$ around the central value $\omega_{\text{pump}}$. We set
$t_h$ and $t_h^{-1}$ as energy and time units.

We now give a short overview of the equilibrium properties (with
$A=0$) of the model, related to our work. We set $U=10$ and vary
$V$. With increasing $V$ the system undergoes around $U\approx 2V$ a
first-order phase transition directly between SDW and CDW phases.
Choosing rather large $U$ we avoid the intermediate bond-order-wave
phase that separates SDW and CDW at smaller
$U$~\cite{Nakamura:2000gk,Tsuchiizu:2002eg,Ejima:2007go}. In both
phases charge excitations are gapped. In the Mott insulating, i.e.,
SDW phase, gapless spin excitations exist and the system displays no
charge order. In contrast, a charge-density wave is characteristic for
the CDW phase while spin excitations are gapped.

In order to solve the time-dependent Hamiltonian $H(t)$, starting from
the Schr\"odinger equation $i\partial\psi(t)/\partial t=H(t)\psi(t)$,
we employ the time-dependent Lanczos method, which is originally
described in Ref.~\cite{park:5870} and later applied and analyzed in
more detail in Ref.~\cite{Mohankumar2006473}, followed by its
applications in nonequilibrium dynamics of strongly correlated systems
in Ref.~\cite{Prelovsek2011.5931p}. The basic idea is that we
approximate the time evolution of $\ket{\psi(t)}$ by a step-vise
change of time $t$ in small increments $\delta t$. At each step, the
Lanczos basis with dimension $M$ is generated resulting in the time
evolution
\begin{equation}
\ket{\psi(t+\delta t)}\simeq e^{-iH(t)\delta t}\ket{\psi(t)}
\simeq\sum_{l=1}^M e^{-i\epsilon_l\delta t}\ket{\phi_l}\braket{\phi_l}{\psi(t)},
\label{eq:3}
\end{equation}
where $\epsilon_l$ and $\ket{\phi_l}$, respectively, are eigenvalues
and eigenvectors of the tridiagonal matrix generated in Lanczos
iteration. We have confirmed that in our calculation, generally $M=30$
provides adequate accuracy when $\delta t\lesssim 0.1$. For even
smaller $\delta t$, $M$ can be smaller.

In the succeeding numerical calculations, we employ periodic boundary
conditions and set $t_d=5$ and $t_0=12.5$. To investigate the time
evolution of the charge order, we define the charge-charge correlation
as
\begin{equation}
C(j;t)=\frac{(-1)^j}{L}\sum_{i=0}^{L-1}
\eval{\psi(t)}{(n_{i+j}-1)(n_i-1)}{\psi(t)},
\label{eq:6}
\end{equation}
where $L$ is the lattice size. 

\begin{figure}
\includegraphics[width=0.45\textwidth]{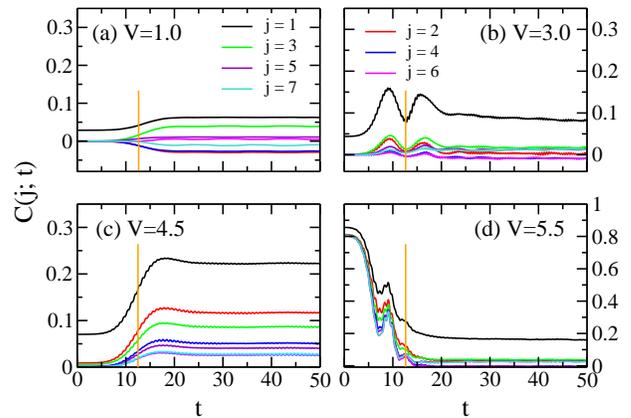}
\caption{(color online) Time dependence of the charge-charge
  correlations as functions of distance (labeled by $j$) for $14$-site
  lattice. The laser pulse with Gaussian magnitude modulation reaches
  its full strength at $t=12.5$, as indicated by solid lines. Without
  exception, the pumping frequencies are set to match the resonance
  peaks of the optical absorption spectrum. Parameters: $U=10$,
  $\delta t=0.02$, $M=100$. (a) $V=1.0$, $\omega_{\text{pump}}=7.1$,
  $A_0=0.10$; (b) $V=3.0$, $\omega_{\text{pump}}=6.1$, $A_0=0.30$; (c)
  $V=4.5$, $\omega_{\text{pump}}=4.0$, $A_0=0.07$; (d) $V=5.5$,
  $\omega_{\text{pump}}=4.1$, $A_0=0.60$.}
\label{fig:1}
\end{figure}

Figure \ref{fig:1} shows the results of charge-charge correlation
functions $C(j;t)$ for the system with $14$ lattice sites, where the
largest distance between two sites is $7$. In order that the photon
energy can be effectively transmitted to the electrons, we match the
pumping frequency $\omega_{\text{pump}}$ with the first optical
absorption peak. Figure~\ref{fig:1}(a) to \ref{fig:1}(c) show the
results of the system approaching the phase boundary from the SDW
side, as the nearest-neighbor interaction $V$ increases from $1.0$ to
$4.5$ (while keeping fixed $U=10$). For $V=1.0$ and $3.0$ the response
of the system shows only a slight increase of mostly short-distance
correlations during and after the pulse. In contrast, at $V=4.5$ the
system displays clear tendency towards a sustainable charge order
enhancement that emerges as the external pulse reaches its maximum and
remains nearly constant after the pulse. By performing a
time-dependent density-matrix renormalization group method under open
boundary conditions on larger system size, up to 30, similar behavior
can be observed (not shown here). We note that the length scale with
which prominent charge order enhancement can be identified in the case
of $V=4.5$ is confined within distance less than ten, in the units of
lattice constant. The opposite effect is found when starting from the
CDW side of the phase diagram, at $V=5.5$ [Fig.~\ref{fig:1}(d)], where
after the pulse, the charge order is substantially diminished, along
with a limited recovery of SDW order.

\begin{figure}
\includegraphics[width=0.45\textwidth]{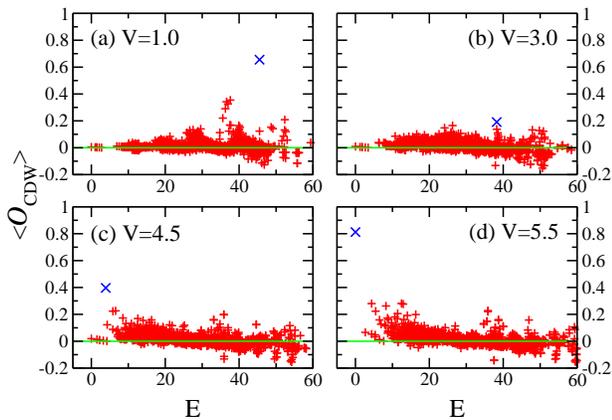}
\caption{(color online) The expectations of CDW order parameter of
  eigenstates for $10$-site lattice with $V=1.0$, $3.0$, $4.5$, and
  $5.5$. The energy $E$ is measured from the ground state. Only the
  data of states with $P_{\text{tot}}=S_{\text{tot}}=0$ are shown, up
  to $E=60$. Note that the largest value of
  $\vev{\mathcal{O}_{\text{CDW}}}$ in (a) through (d) is marked by a
  cross.}
\label{fig:2}
\end{figure}

For deeper understanding of the photoinduced charge order enhancement
we note that the quantum phase transition between SDW-CDW phases is
driven by the competition between the energy cost for doublon
generation and the energy gain due to the attraction between
doublon-holon pairs. The transition point is roughly $V\approx
U/2$. In the SDW phase, we expect charge-order favorite states to
proliferate in the low energy regime as the transition point is
approached, which makes them likely candidates to be picked up by the
laser pulse. This is in contrast to the case when the system is far
from the transition point where CDW states represent high-energy
states. For a more quantitative analysis, we compare the spectra of
systems with different values of $V$ in the time-independent
Hamiltonian, i.e., Eq.~(\ref{eq:1}) but $A(t)=0$. In order to describe
the CDW order with finite spatial extension, we define
\begin{equation}
\mathcal{O}_{\text{CDW}}=\frac{1}{LL_c}\sum_{i=0}^{L-1}\sum_{j=1}^{L_c}
(-1)^j\left(n_{i+j}-1\right)\left(n_i-1\right).
\label{eq:4}
\end{equation}
Here, $L_c$ is introduced as a cut off parameter for the correlation
length. The expectation values of the order parameter
$\vev{\mathcal{O}_{\text{CDW}}}$ for eigenstates of a smaller 10-site
system with $L_c=5$ are produced in Fig.~\ref{fig:2}. Comparison
between different excited states shows that we can distinguish states
with large values of the CDW order. Note that despite the external
laser pulse, the total momentum $P_{\text{tot}}$ and the total spin
$S_{\text{tot}}$ remain good quantum values throughout the time
evolution. For the ground state of a $10$-site lattice at half
filling, we have $P_{\text{tot}}= S_{\text{tot}}=0$. Accordingly, only
results of the eigenstates with the same values of $P_{\text{tot}}$
and $S_{\text{tot}}$ are displayed.

In Fig.~\ref{fig:2}, we find that in the case of small $V=1.0$, the
eigenstates with predominant CDW features are positioned in the high
energy regime, located around $E\sim 45$ above the ground state
energy. With the increase of $V$ the states with large values of
$\langle \mathcal{O}_{\text{CDW}}\rangle $ move towards the lower part
of the energy spectra.  At $V=5.5$ the eigenstate with
$\vev{\mathcal{O}_{\text{CDW}}}\approx 0.65$ turns to be the
ground state [Fig.~\ref{fig:2}(d)]. Keeping this picture in mind, it
becomes more plausible that a well-tuned laser pulse may trigger the
enhancement of the charge order on the SDW side of the phase
diagram. Our numerical calculations suggest that the necessary
precondition for such enhancement is the proximity of the system to
the phase boundary as well as matching conserved quantum numbers
between the SDW and CDW phases.

\begin{figure}
\begin{tabular}{cc}
\includegraphics[width=0.23\textwidth]{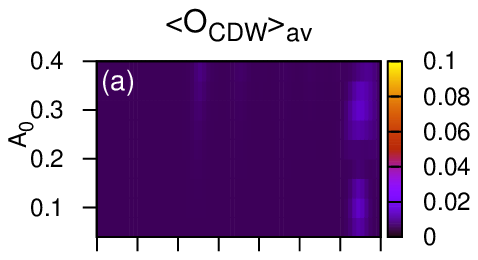}&
\includegraphics[width=0.23\textwidth]{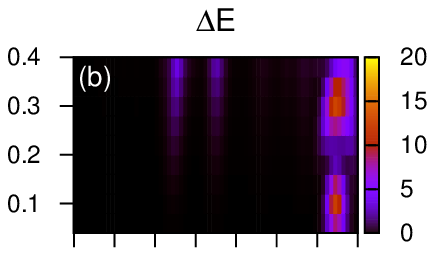}\\
\end{tabular}
\vskip -0.5in
\begin{tabular}{cc}
\includegraphics[width=0.23\textwidth]{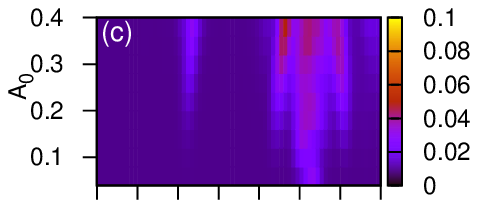}&
\includegraphics[width=0.23\textwidth]{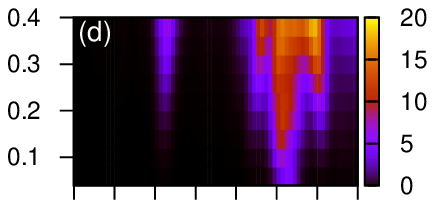}\\
\end{tabular}
\vskip -0.5in
\begin{tabular}{cc}
\includegraphics[width=0.23\textwidth]{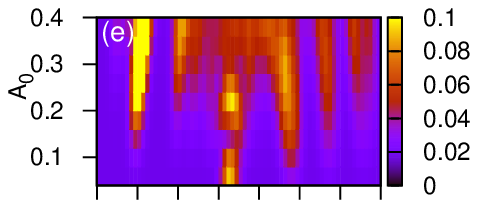}&
\includegraphics[width=0.23\textwidth]{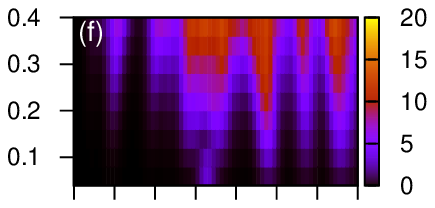}\\
\end{tabular}
\vskip -0.5in
\begin{tabular}{cc}
\includegraphics[width=0.23\textwidth]{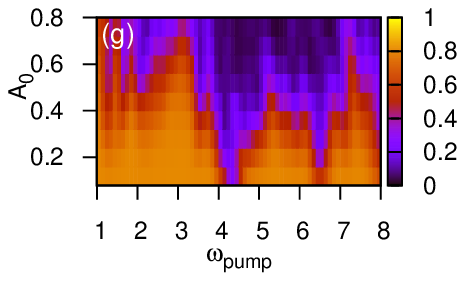}&
\includegraphics[width=0.23\textwidth]{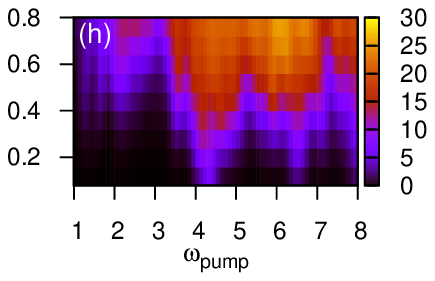}\\
\end{tabular}
\caption{(color online) Contour plots of final time-evolution results
  of CDW order $\vev{\mathcal{O}_{\text{CDW}}}_{\text{av}}$ (first
  column) and energy increase $\Delta E$ (second column) as functions
  of $\omega_{\text{pump}}$ and $A_0$, obtained by averaging on the
  last $50$ time steps (time length $\Delta t=5$), for $10$-site
  lattice. Here, we take $\delta t=0.1$, $M=30$. (a), (b) $V=1.0$;
  (c), (d) $V=3.0$; (e), (f) $V=4.5$; and (g), (h) $V=5.5$.}
\label{fig:3}
\end{figure}

\begin{figure}
\includegraphics[width=0.45\textwidth]{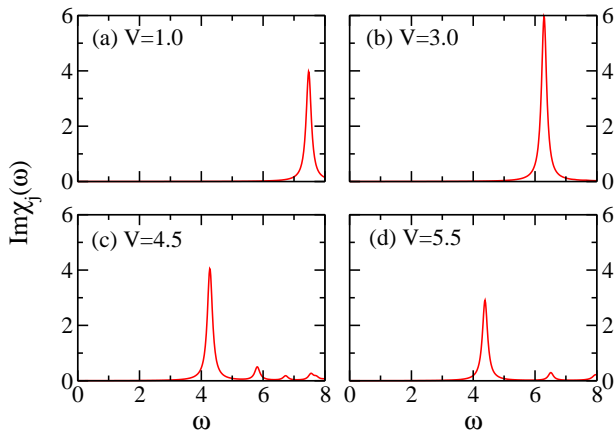}
\caption{(color online) The imaginary part of the dynamical
  current-current correlation function for $10$-site lattice with
  $V=1.0$, $3.0$, $4.5$, and $5.5$, obtained by Lanczos method. The
  $\delta$ function is approached by a Lorentzian with a width of
  $0.1$.}
\label{fig:4}
\end{figure}

We further elaborate on the condition for the emergence of the CDW
order enhancement induced by the laser pulse. To this effect we
perform parameter-sweeping calculations on the $10$-site lattice. We
sweep the pumping frequency $\omega_{\text{pump}}$ and laser intensity
$A_0$ and carry out the time evolutions up to $t=52.5$.  For a given
pair of $\omega_{\text{pump}}$ and $A_0$, we then calculate the
expectation of CDW order $\vev{\mathcal{O}_{\text{CDW}}}_{\text{av}}$,
and the energy increase measured from the ground state $\Delta E$, by
averaging on the last $50$ time steps (corresponding to time length
$\Delta t=5$). Contour plots of
$\vev{\mathcal{O}_{\text{CDW}}}_{\text{av}}$ and $\Delta E$ are shown
in Fig.~\ref{fig:3}. To facilitate the analysis we present in
Fig.~\ref{fig:4} the corresponding optical spectra, obtained from the
imaginary part of the dynamical current-current correlation function:
\begin{eqnarray}
\text{Im}\chi_j(\omega)=\frac{1}{L}\sum_n\abs{\eval{n}{\hat{j}}{0}}^2
\delta\left(\omega-E_n+E_0\right),
\label{eq:8}
\end{eqnarray}
where $\ket{0}$ and $\ket{n}$ represent the ground state and excited
states with energy $E_0$ and $E_n$, respectively. The current operator
$\hat{j}$ defined as
$\hat{j}=-i\sum_{i,\sigma}\left(c_{i,\sigma}^{\dagger}c_{i+1,\sigma}
-\text{H.c.}\right)$.

In Fig.~\ref{fig:3}, we notice the emergence of similar patterns
between $\vev{\mathcal{O}_{\text{CDW}}}_{\text{av}}$ (left) and
$\Delta E$ (right) in each row panel. When starting from the SDW side
[Figs.~\ref{fig:3}(a)-\ref{fig:3}(f)], in the case of moderate values
of $A_0$, the contour plots representing the CDW enhancement and
energy increase roughly match. On the other hand, starting from the
CDW ground state, when $V=5.5$ [Figs.~\ref{fig:3}(g) and
\ref{fig:3}(h)], the effect is just the opposite: the energy increase
occurs along the destruction of charge order. Furthermore, the
positions of the enhanced energy stripes largely coincide with the
resonance-peak positions of optical absorption, as presented in
Fig.~\ref{fig:4}, except for some cases in the large $A_0$ region,
which will be discussed later. From these observations, we can
conclude that at moderate values of $A_0$, in order to obtain enhanced
$\vev{\mathcal{O}_{\text{CDW}}}_{\text{av}}$, the incoming photon
frequency should be tuned close to absorption windows, that match
resonance peaks of the optical spectrum.

Let us investigate the case of $V=4.5$ in more detail. The first
optical peak is located at $\omega\simeq 4.3$
[Fig.~\ref{fig:4}(c)]. States with enhanced CDW order are located
around the same energy (the highest one with energy $3.9$), as shown
in Fig.~\ref{fig:2}(c). Not surprisingly, a sustainable enhancement of
$\vev{\mathcal{O}_{\text{CDW}}}_{\text{av}}$ can be found with
$\omega_{\text{pump}}\simeq 4.3$ when $A_0\sim 0.08$
[Fig.~\ref{fig:3}(e)]. On the other hand, systems with $V=1.0$, and
$3.0$ [Figs.~\ref{fig:3}(a) and~\ref{fig:3}(c)] display no, or at most
a very slight, increase of
$\vev{\mathcal{O}_{\text{CDW}}}_{\text{av}}$ around the resonance
frequencies, which match the optical frequency peaks,
i.e., $\omega\simeq 7.5$ and $6.3$, respectively [Figs.~\ref{fig:4}(a)
and \ref{fig:4}(b)]. We thus propose two conditions for the
observation of the enhancement of
$\vev{\mathcal{O}_{\text{CDW}}}_{\text{av}} $ after the pulse: 1) the
current matrix element $\eval{n}{\hat{j}}{0}$ in Eq.~(\ref{eq:8})
should be sizable and 2) $\eval{n}{\mathcal{O}_{\text{CDW}}}{n}$
should be large as well.

Further detailed analysis on the $V=4.5$ case shows that at
$\omega_{\text{pump}}\approx 4.3$, with $A_0$ growing beyond $0.1$,
$\Delta E$ temporally drops, reaches a local minimum and then keeps
increasing [Fig.~\ref{fig:3}(f)]. The CDW signal up to $A_0\sim 0.25$
follows the same trend and then decreases [Fig.~\ref{fig:3}(e)]. This
suggests a third condition to obtain enhancement of
$\vev{\mathcal{O}_{\text{CDW}}}_{\text{av}}$: the pulse should deliver
the optimal energy increase which can be controlled either by $A_0$,
or another parameter $t_d$. In some simulations we have noticed a
temporary increase of CDW order that lasts only up to the midpoint of
the pulse action and then diminishes. The reason is that the system
has passed the CDW enhancement region during the process -- it has
gained excessive energy. Such a case can be found in
Fig.~\ref{fig:1}(b) ($V=3.0$, $A_0=0.30$), where we observe that two
successive enhancements of the CDW signal appear during the pulse and
then die out. The fact that the double-peak structure also emerges in
the corresponding energy profile is consistent with the above
argument. Moreover, during the irradiation, photons can be absorbed or
emitted at different stages following the time evolution.

We also notice the enhancements of the CDW signal with the
accompanying energy increase when $\omega_{\text{pump}}$ is less than
the resonance frequency, such as for $V=4.5$, when
$\omega_{\text{pump}}\approx 2$ [Fig.~\ref{fig:3}(e)]. We speculate
that this effect is generated by the multiphoton, or more precisely,
two-photon process, partly supported by the fact that it appears at
large $A_0$ and the energy increase is close to $4$, the same as what
happens when the $\omega_{\text{pump}}$ is around $4$. The analysis on
this possible multiphoton process in the off-resonant regime is out of
reach of the present Letter~\cite{Oka:2011vw}.

In conclusion, we address the issues of the thermodynamic limit and
feasible experimental realizations of the proposed CDW enhancement. It
is well known that for the 1D extended Hubbard model, excitonic bound
states can be found at the edge of the optical spectrum when $V\ge
2t_h$~\cite{Matsueda:2004ht}. The existence of well-formed excitons is
a preliminary condition of the enhancement of the CDW order parameter
after a laser pumping. With further increase of $V$, we can expect
that at the edge of the absorption spectrum, charge-order states
multiply and gradually dominate. They can thus be captured by a laser
pulse with well tuned frequency and strength, as proposed by our
numerical simulations.

Although it is quite difficult to find a quasi-1D Mott insulator with
proper $U$ and $V$ values that position the system near the phase
boundary, there are other mechanisms which can take up the role of
$V$. When additional on-site Holstein phonons are taken into account,
it has been shown that the system can be driven from SDW to CDW phase
by electron-phonon
interactions~\cite{Ejima:2010bl,Tezuka:2007}. Representative materials
near the phase boundary with substantial electron-phonon interaction
can be found in tetracyanoquinodimethane (TCNQ)
series~\cite{kumar:234304}. Another way to approach the phase boundary
can be chemical substitutions, as in halogen-bridged transition metal
compounds~\cite{Yamashita:1999bt}. Moreover, a transient phase
transition from CDW to Mott or metallic phases has already been
observed~\cite{Kimura:2009by}. These might prove to be useful in the
search of proper candidates to observe the proposed effect. As an
alternative to electronic systems, CDW enhancement can possibly be
realized in cold atoms~\cite{Lewenstein:2007hr}.

Finally, we would like to emphasize that, while our work has
concentrated on a particular CDW order enhancement induced by a laser
pulse, the proposed mechanism may be effective in other cases when
observing the response of correlated systems to an external
perturbation near a phase boundary such as enhanced antiferromagnetic
correlations, stripes, or possibly even superconducting fluctuations.


This work was also supported by the Strategic Programs for Innovative
Research (SPIRE), the Computational Materials Science Initiative
(CMSI), the global COE program "Next Generation Physics, Spun from
Universality and Emergence" from MEXT, the Yukawa International
Program for Quark-Hadron Sciences at YITP, Kyoto University, and
SLO-Japan collaboration project from ARRS and JSPS. T.T. acknowledges
support by the Grant-in-Aid for Scientific Research (Grant
No. 22340097) from MEXT. J.B. acknowledges support by the P1-0044 of
ARRS, and CINT user program, Los Alamos National Laboratory, NM USA.
A part of numerical calculations was performed in the supercomputing
facilities in YITP and ACCMS, Kyoto University, and ISSP in the
University of Tokyo.

\bibliography{photoinduced1D}

\end{document}